\documentclass[12pt]{article}
\usepackage[utf8]{inputenc}
\usepackage{amsmath,amssymb,aas_macros}
\usepackage{graphicx}
\usepackage{color}
\usepackage{cite}
\usepackage{mathrsfs}

\usepackage[colorlinks=True,citecolor=blue]{hyperref}
\usepackage{indentfirst}
\makeatletter
\@addtoreset{equation}{section}
\makeatother

\usepackage{comment}
\usepackage{listings}
\begin{document}

\begin{titlepage}

\begin{center}

\vskip .75in

{\Large \bf Ruling Out Dark Energy Model induced by de-Sitter Regions of Non-Singular Black Holes with Planck2018, DESI BAO, and Union3 Supernovae}

\vskip .75in

{\large
Shintaro K. Hayashi
}

\vskip 0.25in

{\em
Graduate School of Science, Division of Particle and Astrophysical Science, Nagoya University, Furocho, Chikusa-ku, Nagoya, Aichi 464-8602, Japan

}

\begin{abstract}

Dark energy(DE) remains one of the most important subjects in modern cosmology, and its physical origin is still under intensive discussion.
While an astrophysical origin of DE is a highly challenging scenario, black holes stand out the most promising candidates as the astrophysical origin.
In this paper, we explore a new model of DE induced by black holes, in which cosmic accelerated expansion caused by de-Sitter like space-time regions around the non-singular black holes.
It is difficult to examine such a phenomena by measuring black hole mass because the energy density of the cosmological constant is much smaller than the mass density of a black hole near a black hole. On the other hand, this modification becomes dominant on the cosmological scale. Therefore, we focus on the cosmological probes and perform the MCMC analysis using Planck2018+DESI DR2+Supernovae. Since the total amount of DE density depends on the contributions of all black holes, we use the simulated results for the evolution of the number of black holes. As a result, we obtain the best-fitted total chi-squared value, $\chi_\mathrm{total}^2 = 2871.13$ compared to $\Lambda \mathrm{CDM}$ case $\chi_\mathrm{total, \, \Lambda CDM}^2 = 2819.00$, and $\Delta \chi^2 \sim 50$.
We conclude that this $\Delta \chi^2$ is enough large to rule out this model, because the number of parameters is same between this model and $\Lambda \mathrm{CDM}$.

\end{abstract}

\end{center}
\vskip .5in

\end{titlepage}

\section{Introduction}
\label{sec:intro}

In the standard $\Lambda$CDM model, we introduce the cosmological constant $\Lambda$ to explain the cosmic accelerated expansion. The cosmological constant can be considered as the geometrical modification of the general relativity, or the constant energy density components. The physical origin of DE remains unresolved, and various models have been proposed, including modified gravity and quintessence. The measurements of time evolution of DE energy density is the key to reveal the physical origin of DE. Recent observations of baryon acoustic oscillations (BAO) by the Dark Energy Spectroscopic Instrument (DESI) suggest that the dark energy density may vary over time, as modeled by the $w_0w_a$CDM parameterization \cite{DESI:2024mwx, DESI:2025fii}.

Black holes (BHs) have been proposed as a potential source of DE, and a representative model is the "cosmologically coupled black hole (CCBH)" scenario, in which the BH mass evolves as $M \propto a^3$, where $a$ is the scale factor. Because the energy density of non-relativistic matter scales as $\rho \propto a^{-3}$, the CCBH model yields an approximately constant effective energy density on cosmological scales. This raises the intriguing possibility that BHs could contribute to the dark energy component. Moreover, the number density of BHs changes through the star formation and BH formation, so it generates the time dependence of dark energy. In Ref.~\cite{Croker:2024jfg}, the authors demonstrated that the CCBH model can effectively reproduce the $w_0w_a$CDM evolution in the DESI BAO data. They further extended their analysis using Planck CMB+ DESI BAO+ Supernovae data in Ref.~\cite{2025arXiv250420338A}. In these studies, the CCBH framework is interpreted as a type of baryon–dark energy interaction model, with the interaction rate assumed to be proportional to the star formation rate \cite{2022MNRAS.511..616T, 2024MNRAS.529.3563T, 2014ARA&A..52..415M, 2017ApJ...840...39M}. Despite these successes, several astrophysical studies estimating the black hole rest masses suggest that the mass evolution $M \propto a^3$ is disfavored \cite{Lacy:2023kbb, Lei:2023mke, 2023A&A...673L..10A}. Therefore, we consider an alternative model of dark energy sourced by BHs in this paper. 

From a theoretical perspective, the cosmological extension of BH model, such as CCBH model, incorporates black hole geometries into an expanding universe described by the Friedmann–Robertson–Walker (FRW) spacetime \cite{2007PhRvD..76f3510F}. This extension is also relevant to efforts to resolve black hole singularities \cite{2018CQGra..35w5005E, 2023JCAP...11..007C, Gielen:2025ovv, Misyura:2024fho}. According to Ref.~\cite{2018CQGra..35w5005E}, it has been shown that non-singular black holes can be described by the Schwarzschild–de Sitter (SdS) geometry outside the event horizon. This result can be interpreted as indicating that a local de Sitter region with cosmological constant $\Lambda_i$ is enhanced around each BH. The observational constraints that disfavor the CCBH model \cite{Lacy:2023kbb, Lei:2023mke, 2023A&A...673L..10A} originate from small-scale ($\lesssim$ kpc) physical phenomena of individual BH masses. In contrast, in the SdSDE model, $\Lambda_i$ is much smaller than the BH rest mass at these scales, and its effects on the local physics are negligible. Therefore, these constraints do not apply to the SdSDE scenario. Since our universe contains a large number of BHs, the collective contribution of these local de Sitter regions, summed as $\sum \Lambda_i$, may effectively constitute the background cosmological constant. We refer to this concept as the Schwarzschild–de Sitter black hole dark energy (SdSDE) model. The total SdSDE energy density then depends on the cosmic number density of BHs. In previous CCBH studies~\cite{Croker:2024jfg, 2025arXiv250420338A}, the redshift evolution of the BH number density was estimated from the cosmic star formation rate. In this work, we instead adopt the BH mass function obtained from population synthesis simulations that include binary evolution and merger histories \cite{2022ApJ...924...56S, 2022ApJ...934...66S}. This provides an alternative, simulation-based estimate of the BH number density evolution.

In this paper, we model the evolution of the SdSDE energy density using the simulated BH mass function in Ref.~\cite{2022ApJ...924...56S}, and we test the SdSDE scenario using Planck 2018 CMB data, DESI BAO DR2, and Union3 supernova. We also examine possible astrophysical effects of SdS region under the constraints imposed by the combination of the cosmological dataset.

In Sec.~\ref{subsec:ccbh}, we introduce the specific SdSDE model adopted in this work. Since its energy density depends on the BH density, we describe the mass function and its evolution in Sec.~\ref{subsec:rhobh}, following Ref.~\cite{2022ApJ...924...56S}. The constraints on cosmological parameters obtained via Markov Chain Monte Carlo (MCMC) analysis are presented in Sec.~\ref{sec:results}. Finally, Sec.~\ref{sec:summary} provides discussion and concluding remarks.

\section{Methods}
\label{sec:method}

\subsection{Dark Energy induced by Non-Singular SdS Black Holes}
\label{subsec:ccbh}
In this paper, we start with the Schwarzschild–de Sitter (SdS) black hole, which may describe the exterior geometry of a non-singular black hole outside the event horizon~\cite{2018CQGra..35w5005E}. Its line element is given by
\begin{equation}
    ds^2 = - f(r) \, dt^2 + f(r)^{-1} \, dr^2 + r^2 \left( d \theta^2 + \sin^2 \theta \, d \phi^2 \right),
\end{equation}
where
\begin{equation}
    f(r) = 1 - \frac{2GM}{r} - \frac{\Lambda}{3} r^2, \\ 
\end{equation}
and $M$ is the conventional mass of BH, originating from the matter content involved in the BH formation.

In the SdSDE model, the cosmological constant is not regarded as a universal constant but is instead sourced by the mass distribution of black holes. In this framework, the effective mass profile of a black hole can be written as
\begin{equation}
\label{mass_eff}
    M_\mathrm{eff}(r) = M + C r^3,
\end{equation}
and
\begin{equation}
    f(r) = 1 - \frac{2GM_\mathrm{eff}\left(r\right)}{r},
\end{equation}
where $C$ is a constant with dimensions of energy density. Physically, this implies that the cosmological term contributes a position-dependent mass that becomes dominant at large distances.

In the limit $r \to \infty$, the additional mass term dominates, leading to a constant mass density associated with the black hole. Since the volume enclosed by a comoving radius evolves as $a^3$, this suggests that the effective mass of a CCBH evolves as $M_\mathrm{eff} \propto a^3$ on cosmological scales. Therefore, a collection of such black holes can collectively give rise to an energy density that mimics a dark energy component.

Assuming that the cosmological constant arises from a sum of contributions from individual black holes, we write
\begin{align}
    \rho_\mathrm{tot} &= \sum_i \left(\frac{M_i}{r^3} +  C_i \right) \\
    \label{totbhenergy}
    &\simeq \frac{M_\mathrm{mean}N_\mathrm{BH}}{r^3} + C_\mathrm{mean} \times N_\mathrm{BH},
\end{align}
where $C_i$ denotes the contribution to the effective cosmological constant of the $i$-th black hole and $N_\mathrm{BH}$ is the number of black holes. Although $M_i$ and $C_i$ may in general depend on the properties of individual black holes, we approximate all contributions by mean values $M_\mathrm{mean}$ and $C_\mathrm{mean}$ corresponding to a representative black hole.

Since the first term of Eq. \ref{totbhenergy} corresponds to the rest-mass energy density of black holes, the dark energy density can then be approximated as
\begin{align}
    \rho_\mathrm{DE} &\simeq C_\mathrm{mean} \, n_\mathrm{BH} \, r^3\\
    &\simeq C_\mathrm{mean} \times M_\mathrm{mean}n_\mathrm{BH} \times \frac{\bar{r}^3}{M_\mathrm{mean}} \, a^3
\end{align}
where $n_\mathrm{BH}$ is the number density of black holes written as $n_\mathrm{BH} \equiv N_\mathrm{BH}/r^3$, and $\bar{r}$ is the comoving coordinates defined by $\bar{r} = r/a$. The part of $M_\mathrm{mean}n_\mathrm{BH}$ can be interpreted as $\rho_\mathrm{tot,BH}$. $\bar{r}^3/M_\mathrm{mean}$ corresponds to the inverse of the comoving mass density of a black hole, so this is constant, and we can define the new constant coefficient $D \equiv C \bar{r}^3 / M_\mathrm{mean}$. Then, the dark energy density becomes
\begin{equation}
\label{rho_ccbh}
    \rho_\mathrm{DE} \simeq D \, \rho_{\mathrm{tot,BH}} \, a^3,
\end{equation}
where $\rho_{\mathrm{tot,BH}}$ is the physical mass density for all BHs. Hereafter, we call this dark energy density as "SdSDE".

Since the region affected by $C$ is expected to expand at the speed of light, it may induce additional inhomogeneities or perturbations in the dark energy sector. However, in this paper, we neglect such perturbative effects and focus only on the background evolution of the SdSDE energy density.

\subsection{Time evolution of BH mass density}
\label{subsec:rhobh}

In this analysis, we adopt the black hole (BH) mass density derived from simulations based on a binary evolution code, as reported in Ref.~\cite{2022ApJ...924...56S}. A convenient analytical fitting function, along with its corresponding best-fit parameters, is used to describe the mass function.

As discussed in Ref.~\cite{2022ApJ...924...56S}, the BH mass function is well approximated by a combination of a Schechter function and a Gaussian component \cite{1990MNRAS.242..318S}, given by:
\begin{equation}
    \label{schechtergaussian}
    \frac{dN}{dV d \log m_\bullet} \simeq \mathcal{N} \left( \frac{m_\bullet}{\mathcal{M}_\bullet} \right)^{1 - \alpha} e^{- m_\bullet / \mathcal{M}_\bullet}
    + \mathcal{N}_\mathrm{G} \frac{1}{\sqrt{2\pi \sigma_\mathrm{G}^2}} \exp\left[ -\frac{ \left( \log m_\bullet - \log \mathcal{M}_{\bullet, \mathrm{G}} \right)^2 }{2\sigma_\mathrm{G}^2} \right],
\end{equation}
where $N$ is the number of black holes, $V$ is the comoving volume, and $m_\bullet$ is the black hole mass. The parameters $\mathcal{N}$, $\mathcal{M}_\bullet$, $\alpha$, $\mathcal{N}_\mathrm{G}$, $\mathcal{M}_{\bullet,\mathrm{G}}$, and $\sigma_\mathrm{G}$ are redshift-dependent fitting coefficients.

\begin{figure}
    \centering
    \includegraphics[width=1.0\linewidth]{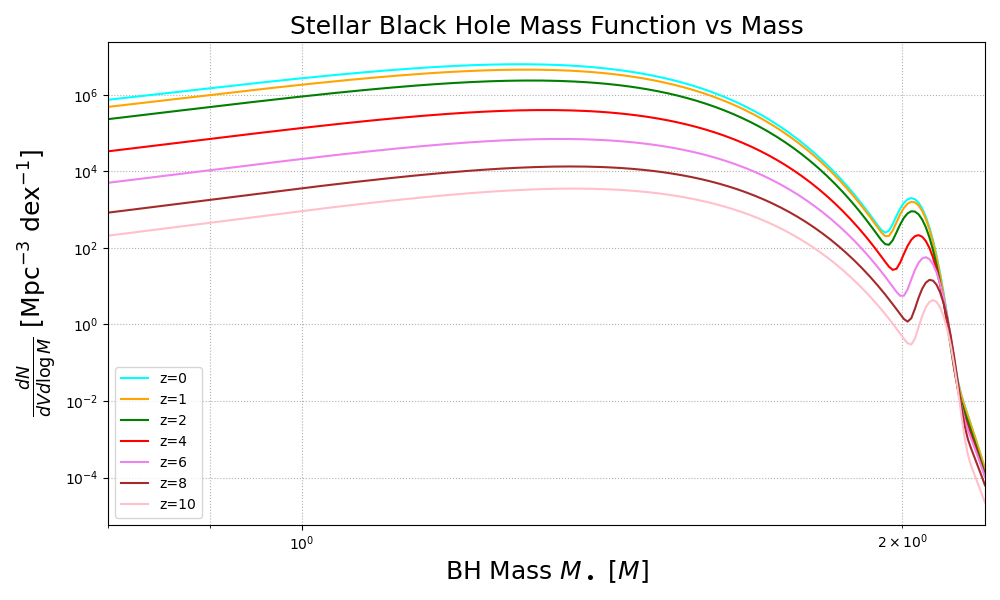}
    \caption{Comoving BH mass functions at redshifts $z = 0$, $1$, $2$, $4$, $6$, $8$, and $10$, using the fitting function in Eq.~\ref{schechtergaussian} with the best-fit parameters from Table~\ref{tab_params2}.}
    \label{fig:massfunc}
\end{figure}

The best-fit parameters at various redshifts, obtained in Ref.~\cite{2022ApJ...924...56S}, are listed in Table~\ref{tab_params2}. Using these results, the comoving BH mass function at several redshifts is shown in Fig.~\ref{fig:massfunc}.

\begin{table}[ht]
    \centering
    \renewcommand{\arraystretch}{1.5}
    \begin{tabular}{|c|c|c|c|c|c|c|}
    \hline
    $z$ & $\log \mathcal{N} \, (\mathrm{Mpc}^{-3})$ & $\log \mathcal{M}_\bullet \, (M_\odot)$ & $\alpha$ & $\log \mathcal{N}_\mathrm{G} \, (\mathrm{Mpc}^{-3})$ & $\log \mathcal{M}_{\bullet,\mathrm{G}} \, (M_\odot)$ & $\sigma_\mathrm{G}$ \\ \hline \hline
    $0$ & $5.623$ & $0.607$ & $-3.781$ & $2.413$ & $2.021$ & $0.052$ \\ \hline
    $1$ & $5.429$ & $0.609$ & $-3.859$ & $2.309$ & $2.023$ & $0.051$ \\ \hline
    $2$ & $5.107$ & $0.612$ & $-3.914$ & $2.064$ & $2.024$ & $0.051$ \\ \hline
    $4$ & $4.344$ & $0.634$ & $-3.902$ & $1.419$ & $2.037$ & $0.049$ \\ \hline
    $6$ & $3.614$ & $0.659$ & $-3.866$ & $0.806$ & $2.054$ & $0.045$ \\ \hline
    $8$ & $2.894$ & $0.676$ & $-3.868$ & $0.197$ & $2.066$ & $0.043$ \\ \hline
    $10$ & $2.305$ & $0.680$ & $-3.884$ & $-0.344$ & $2.072$ & $0.042$ \\ \hline
    \end{tabular}
    \caption{Best-fit parameters for the Schechter+Gaussian BH mass function at several redshifts, from Ref.~\cite{2022ApJ...924...56S}.}
    \label{tab_params2}
\end{table}

\begin{figure}
    \centering
    \includegraphics[width=1.0\linewidth]{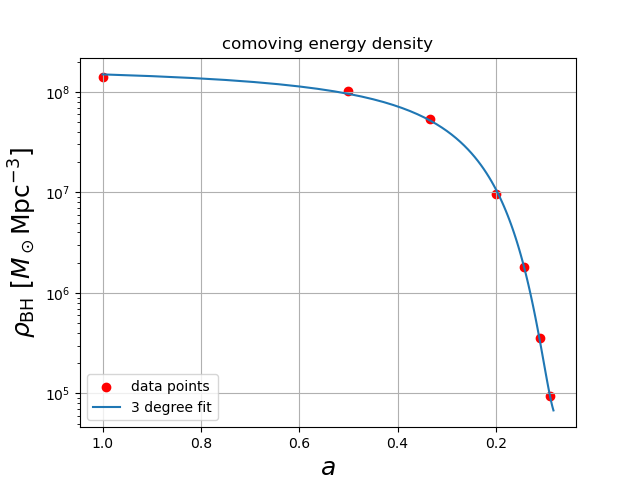}
    \caption{Comoving BH mass density as a function of redshift. Red points are computed from Eq.~\ref{rho_from_massfunc} using the mass functions in Fig.~\ref{fig:massfunc}. The solid blue line shows the polynomial fit described by Eq.~\ref{rho_polynomial}.}
    \label{fig:BHrho}
\end{figure}

The corresponding BH mass density can be computed from the mass function via:
\begin{equation}
\label{rho_from_massfunc}
    \rho_\mathrm{BH}(z) = \int d\log m_\bullet \, m_\bullet \, \frac{dN}{dV d\log m_\bullet}(m_\bullet | z).
\end{equation}
The resulting redshift evolution of the BH mass density is shown as red points in Fig.~\ref{fig:BHrho}.

For the estimation of the cosmological parameters, a continuous function of $\rho_\mathrm{BH}(z)$ is required. We therefore perform a polynomial fit of the form:
\begin{equation}
\label{rho_polynomial}
    \log \rho_\mathrm{BH}(z) = a z^3 + b z^2 + c z + d,
\end{equation}
where the current BH mass density is given by $\rho_{0\mathrm{BH}} = e^d \, [M_\odot\, \mathrm{Mpc}^{-3}]$.

To select the optimal degree of the polynomial in Eq. \ref{rho_polynomial}, we perform cross-validation. The validation scores are: degree 1: 0.070, degree 2: 0.182, degree 3: 0.065, degree 4: 0.135, and degree 5: 0.325. Based on this evaluation, a degree-3 polynomial provides the best balance between bias and variance. The best-fit parameters are listed in Table~\ref{tab_params3}, and the resulting function is shown as the solid blue line in Fig.~\ref{fig:BHrho}.

\begin{table}[ht]
    \centering
    \renewcommand{\arraystretch}{1.5}
    \begin{tabular}{|c|c|c|c|}
    \hline
    $a$ & $b$ & $c$ & $d$ \\ \hline \hline
    $0.00658$ & $-0.104$ & $-0.348$ & $18.8$ \\ \hline
    \end{tabular}
    \caption{Best-fit coefficients for the polynomial fit to $\log \rho_\mathrm{BH}(z)$.}
    \label{tab_params3}
\end{table}

\subsection{Implementation into \texttt{Cobaya}}
\label{subsec:imple}
We use the \texttt{cobaya} package \cite{2021JCAP...05..057T, 2019ascl.soft10019T} to estimate the cosmological parameters and implement the evolution of the SdSDE energy density into \texttt{CAMB} by modifying the dark energy component at the background level. In this study, we consider only the background evolution and neglect the perturbations of the dark energy component.

According to Eqs.~\ref{rho_ccbh} and \ref{rho_polynomial}, the energy density of the dark energy component is modified as:
\begin{align}
    \rho_\mathrm{DE}(z) &= D \rho_\mathrm{tot,BH}(z) \, a^3 \\
    &= D \times \frac{1}{a^3} \rho_{0\mathrm{BH}} \exp[az^3 + bz^2 + cz] \times a^3 \\
    &= \rho_{0\mathrm{DE}} \, \exp[az^3 + bz^2 + cz],
\end{align}
where $\rho_{0\mathrm{DE}}$ is the current energy density of SdSDE. It is determined by the other cosmological parameters as
\[
\rho_{0\mathrm{DE}} = \Omega_\Lambda \frac{3H_0^2}{8\pi G} = \left(1 - \Omega_m\right)\frac{3H_0^2}{8\pi G}.
\]
From Eq.~\ref{mass_eff}, the total baryon energy density is conserved in this model, since the baryonic mass incorporated into black hole formation is conserved as the conventional black hole mass. Therefore, the number of free parameters in the SdSDE model remains the same as in the standard $\Lambda$CDM model.

For the Markov Chain Monte Carlo (MCMC) analysis, we adopt flat priors for the cosmological parameters, as listed in Table~\ref{tab_priors}. We use the Planck 2018 low-$\ell$ and high-$\ell$ temperature and polarization data \cite{Planck:2019nip, Planck:2018vyg}, and also include CMB lensing data \cite{Planck:2018lbu}, DESI DR2 BAO data \cite{DESI:2025fii}, and Union3 supernova data \cite{2025ApJ...986..231R}.

MCMC sampling is terminated when convergence is achieved according to the Gelman–Rubin criterion \cite{10.1214/ss/1177011136}, with condition $R-1 < 0.01$ applied to all runs.

\begin{table}[ht]
    \centering
    \renewcommand{\arraystretch}{1.5}
    \begin{tabular}{|c|c|}
    \hline
    Parameter & Prior Range \\ \hline
    $\Omega_{\rm b} h^2$ & [0.05, 0.1] \\
    $\Omega_{\rm c} h^2$ & [0.001, 0.99] \\
    $\tau$ & [0.01, 0.8] \\
    $100\Theta_{\rm s}$ & [0.5, 10] \\
    $n_{\rm s}$ & [0.7, 1.2] \\
    $\log_{10} A_{\rm s}$ & [2, 5] \\
    \hline
    \end{tabular}
    \caption{Flat prior ranges for the cosmological parameters used in the MCMC analysis.}
    \label{tab_priors}
\end{table}

\section{Results and Discussion}
\label{sec:results}
\begin{figure}
\centering
\includegraphics[width=15cm]{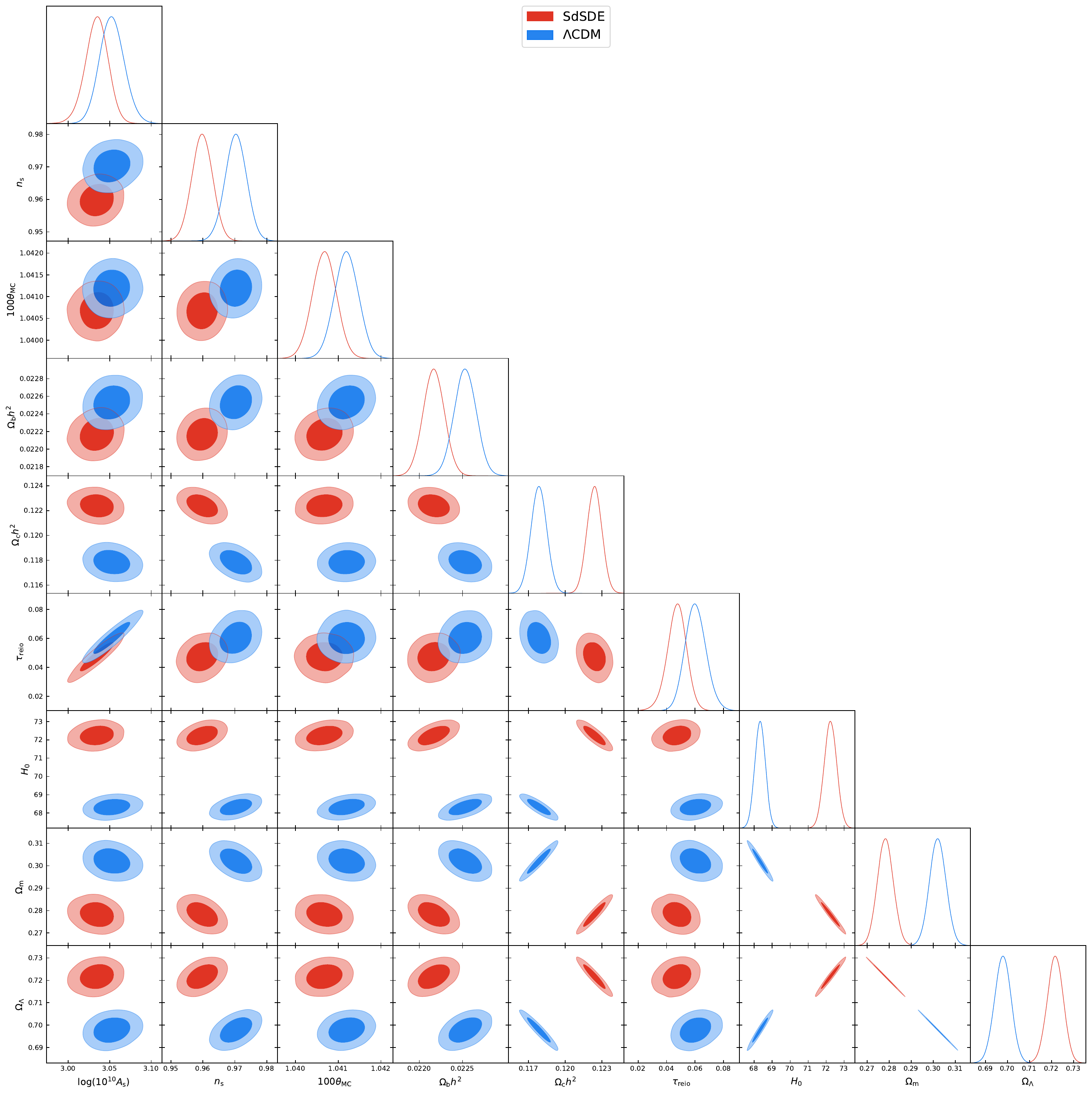}
\caption{Constraints on the cosmological parameters from MCMC analysis. Red and blue contours represent the SdSDE and $\Lambda$CDM models, respectively.}
\label{fig:mcmc_cosmo1}
\end{figure}

\begin{figure}
\centering
\includegraphics[width=15cm]{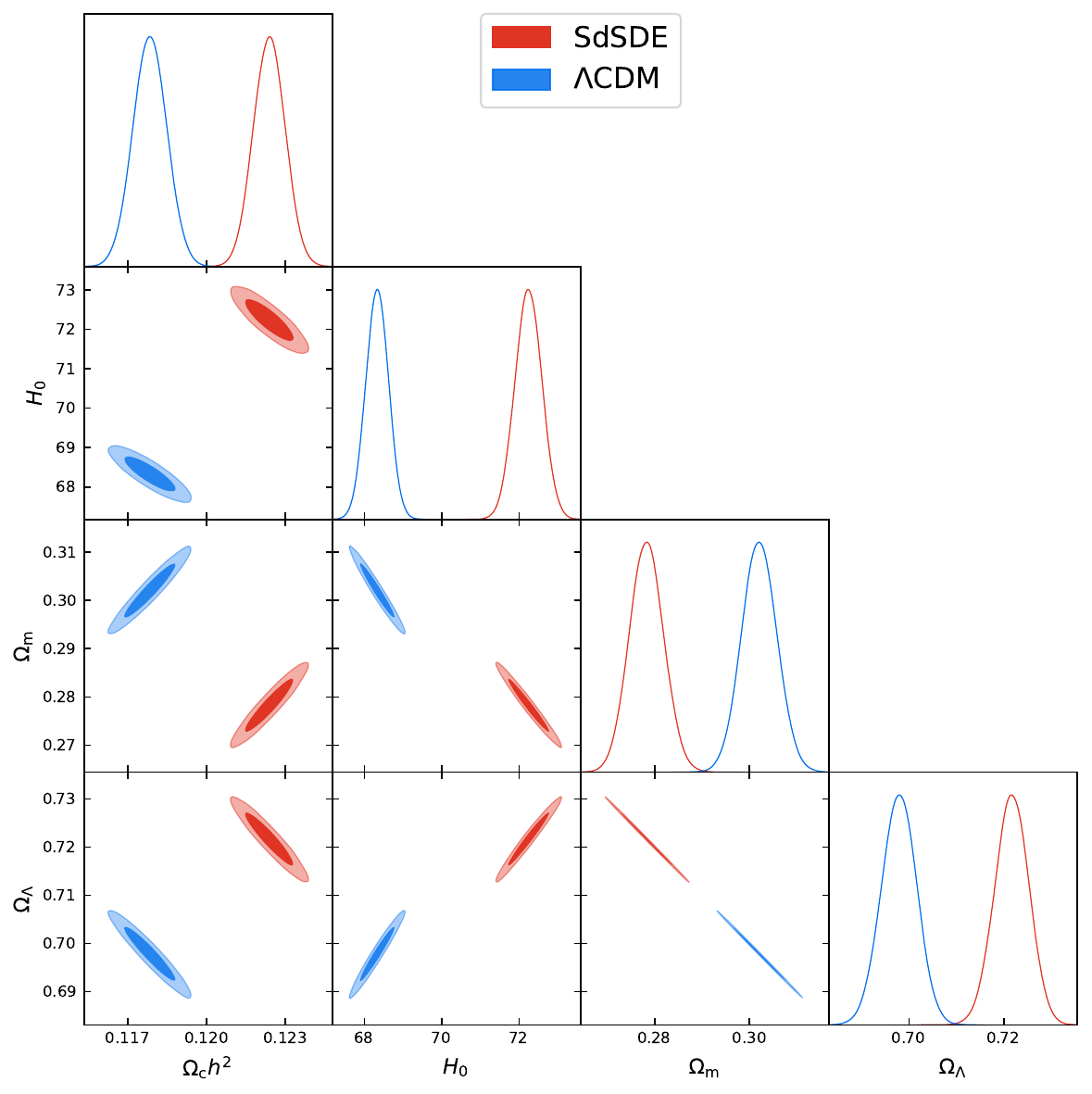}
\caption{Constraints in the $H_0$–$\Omega_\mathrm{m}$, $H_0$–$\Omega_\Lambda$, and $H_0$–$\Omega_\mathrm{c} h^2$ planes. The SdSDE model shows larger values of $H_0$, $\Omega_\Lambda$, and $\Omega_\mathrm{c} h^2$, and a smaller value of $\Omega_\mathrm{m}$ compared to the $\Lambda$CDM model.}
\label{fig:mcmc_cosmo2}
\end{figure}

First, we present the triangle plot of the cosmological parameters obtained from the MCMC analysis in Figure \ref{fig:mcmc_cosmo1}. A selection of key parameters, $H_0$, $\Omega_\mathrm{m}$, $\Omega_\Lambda$, and $\Omega_\mathrm{c} h^2$, are highlighted in Figure \ref{fig:mcmc_cosmo2}. The SdSDE model yields higher $H_0$, $\Omega_\Lambda$, and $\Omega_\mathrm{c} h^2$ and a lower $\Omega_\mathrm{m}$ compared to the $\Lambda$CDM model. This is due to the rapid growth of the dark energy component in the SdSDE model around the scale factor $a \sim 0.2$–0.4. To compensate for the reduced total energy density at high redshifts, the present total energy density increases, leading to higher values of $\Omega_\Lambda$, $H_0$, and $\Omega_\mathrm{c} h^2$. In contrast, $\Omega_\mathrm{m}$ decreases because it is given by $1 - \Omega_\Lambda$.

\begin{table}[ht]
    \centering
    \renewcommand{\arraystretch}{1.5}
    \begin{tabular}{|c|c|c|}
    \hline
    Parameters & SdSDE & $\Lambda \mathrm{CDM}$ \\ \hline
    $H_0$ & $72.24 \left( 72.32 \right) ^{+0.34}_{-0.35}$ & $68.33 \left( 68.21 \right) ^{+0.29}_{-0.29}$ \\
    $\Omega_{\Lambda}$ & $0.7217 \left( 0.7220 \right) ^{+0.0036}_{-0.0036}$ & $0.6978 \left( 0.6961 \right) ^{+0.0037}_{-0.0037}$ \\
    $\Omega_{\rm m}$ & $0.2783 \left( 0.2779 \right) ^{+0.0036}_{-0.0036}$ & $0.3021 \left( 0.3038 \right) ^{+0.0037}_{-0.0037}$ \\ \hline
    $100\Omega_{\rm b} h^2$ & $2.217 \left( 2.226 \right) ^{+0.012}_{-0.012}$ & $2.253 \left( 2.253 \right) ^{+0.013}_{-0.013}$\\
    $\Omega_{\rm c} h^2$ & $0.1224 \left( 0.1225 \right) ^{+0.0006}_{-0.0006}$ & $0.1178 \left( 0.1182 \right) ^{+0.0006}_{-0.0006}$ \\
    $\tau$ & $0.0471 \left( 0.0470 \right) ^{+0.0070}_{-0.0062}$ & $0.0604 \left( 0.0533 \right) ^{+0.0069}_{-0.0076}$\\
    $100\Theta_{\rm{s}}$ & $1.0407 \left( 1.0408 \right) ^{+0.0003}_{-0.0003}$ & $1.0412 \left( 1.0412 \right) ^{+0.0003}_{-0.0003}$ \\
    $n_{\rm{s}}$ & $0.9598 \left( 0.9591 \right) ^{+0.0032}_{-0.0032}$ & $0.9703 \left( 0.9701 \right) ^{+0.0033}_{-0.0033}$\\
    $\mathrm{log}_{10}A_{\rm{s}}$ & $3.034 \left( 3.034 \right) ^{+0.014}_{-0.013}$ & $3.053 \left( 3.040 \right) ^{+0.014}_{-0.015}$\\
    \hline
    \end{tabular}
    \caption{Mean values and 68\% confidence level for some cosmological parameters in each model obtained from the MCMC analysis using the Planck 2018 data. The best-fit values are shown in the parentheses.}
    \label{tab_results}
\end{table}

Next, we examine the impact of the SdS region parameter $C$. The critical density is calculated as $\rho_\mathrm{crit} = 3 H_0^2 / 8 \pi G$, which yields $\rho_\mathrm{crit} \simeq 1.452 \times 10^{11} \,  \mathrm{M}_\odot / \mathrm{Mpc}^3 $ using the best-fit parameters. The contribution from all SdSDEs is $\Omega_\Lambda \rho_\mathrm{crit} \simeq 1.047 \times 10^{11} \, \mathrm{M}_\odot / \mathrm{Mpc}^3 $. In contrast, the current black hole mass density is given by $\rho_{0\mathrm{BH}} = e^d$, and with $d = 18.8$, we obtain $\rho_{0\mathrm{BH}} \simeq 1.461 \times 10^8 \,  \mathrm{M}_\odot / \mathrm{Mpc}^3 $. Assuming each black hole has mass $M = 10 \, \mathrm{M}\odot$, the implied effects of SdS region per black hole is $10^3 \, \mathrm{M}_\odot$ per $\mathrm{Mpc}^3$, suggesting that each SdSDE contributes far more than its rest mass. However, since the SdSDE contribution arises from a constant density term, its effect becomes negligible at small scales. For example, on $\sim$pc scales typical of binary black hole separations, the contribution of SdS region is only $10^{-15} \, \mathrm{M}_\odot$, effectively undetectable. Thus, gravitational wave observations of binary systems are unlikely to probe the SdSDE effect.

Now we turn to the comparison between the SdSDE and $\Lambda$CDM models. Although the contours in Figure \ref{fig:mcmc_cosmo1} differ, this is expected because the SdSDE parameterization is not a smooth extension of $\Lambda$CDM (such as $w_0 w_a$CDM). Since both models have the same number of free parameters, a direct comparison using total $\chi^2$ is valid.

\begin{figure}
\centering
\includegraphics[width=15cm]{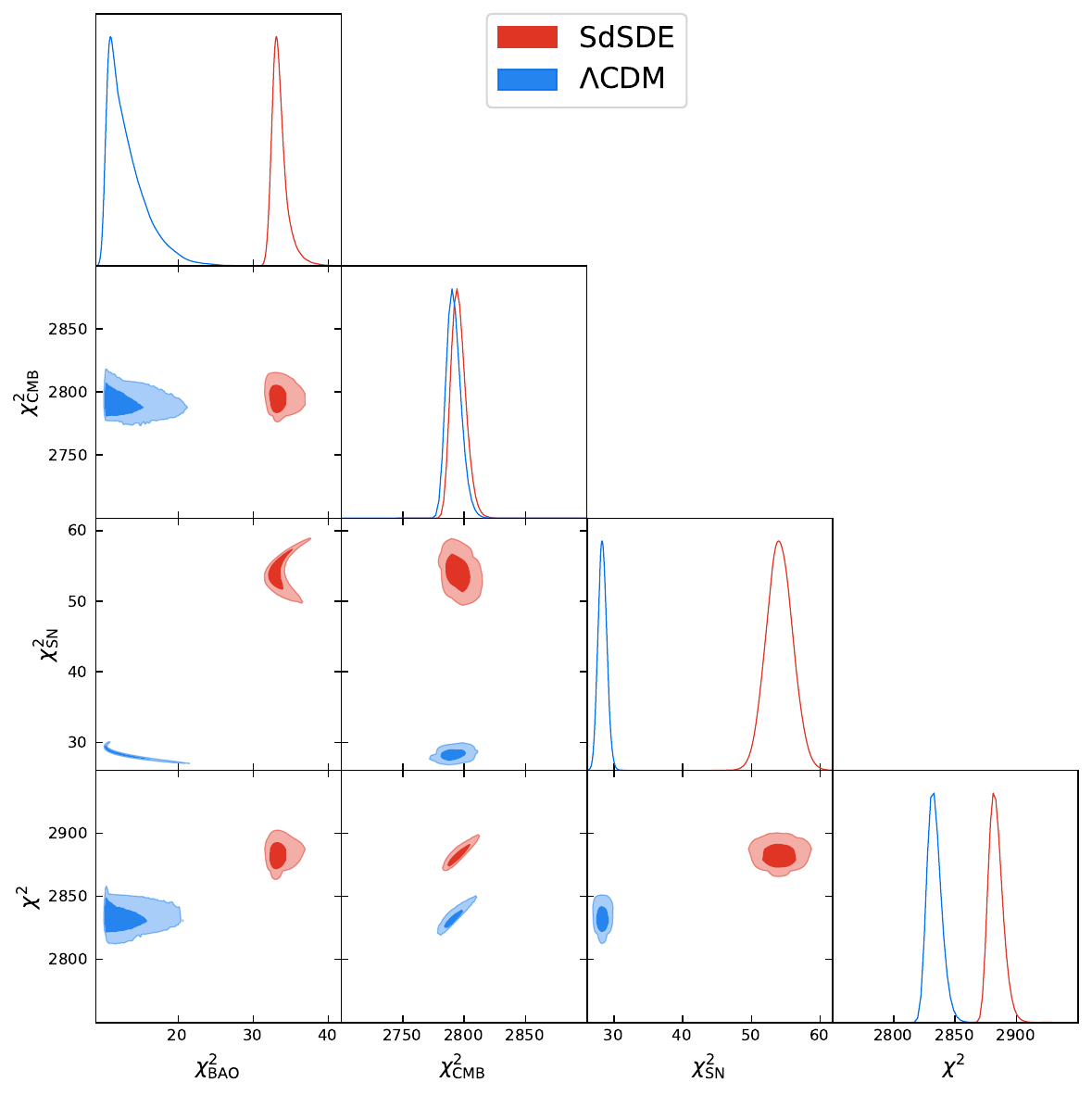}
\caption{Total and component-wise $\chi^2$ values. $\chi^2_\mathrm{CMB}$, $\chi^2_\mathrm{BAO}$, and $\chi^2_\mathrm{SN}$ correspond to Planck+lensing, DESI DR2 BAO, and supernovae, respectively. The SdSDE model exhibits larger $\chi^2$ than $\Lambda$CDM.}
\label{fig:mcmc_chi}
\end{figure}

Figure \ref{fig:mcmc_chi} and Table \ref{tab_params4} present the total and individual $\chi^2$. The total $\chi^2$ is decomposed into $\chi^2_\mathrm{CMB}$, $\chi^2_\mathrm{BAO}$, and $\chi^2_\mathrm{SN}$, corresponding to Planck+lensing, DESI DR2 BAO, and supernova data. While the fits to the CMB are comparable between the two models, the SdSDE model yields significantly worse fits for BAO and supernovae data. 

Since the two models have the same number of free parameters, the difference in the total $\chi^2$ directly reflects their relative statistical performance. We find
\begin{equation}
    \Delta \chi^2 = \chi^2_\mathrm{SdSDE}- \chi^2_{\Lambda \mathrm{CDM}} = 52.13,
\end{equation}
in favor of $\Lambda$CDM. Because the parameter dimensionality is identical, the Akaike Information Criterion(AIC) and Beyesian Information Criterion(BIC) differences reduce to
\begin{equation}
\Delta \mathrm{AIC} = \Delta \mathrm{BIC} = 52.13 ,  
\end{equation}
providing decisive evidence against SdSDE. We therefore conclude that the SdSDE model is strongly disfavored by the combined CMB + BAO + SN data.

\begin{figure}
    \centering
    \includegraphics[width=10cm]{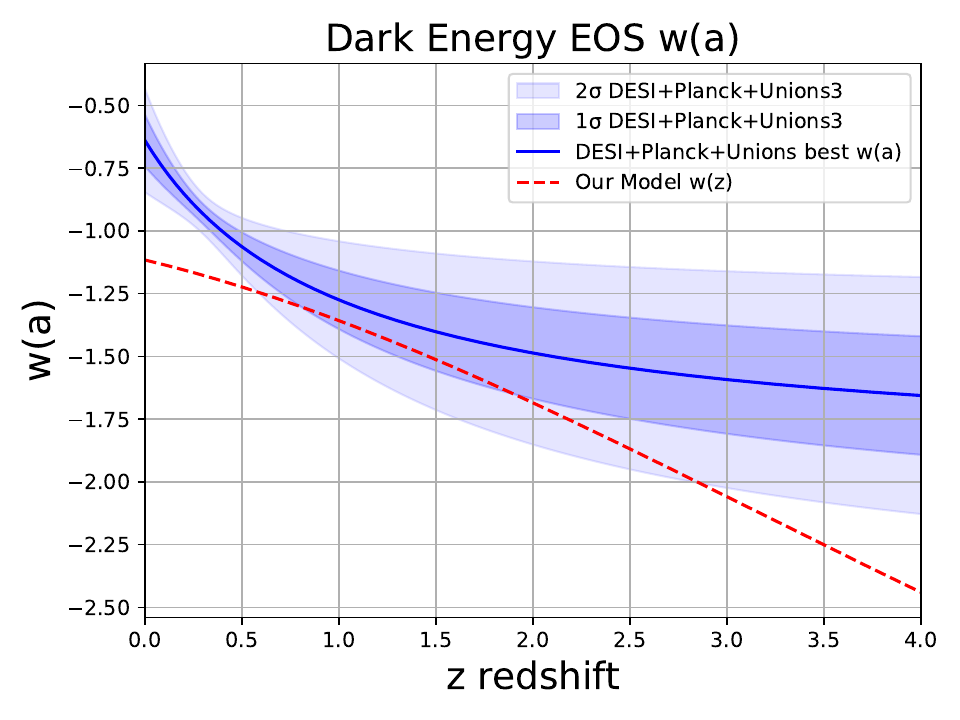}
    \caption{The evolution of the dark energy equation of state with redshift. The red dashed line shows the best-fit $w(a)$ of the SdSDE model. The blue solid line and shaded regions indicate the best fit and the 68\% and 95\% confidence levels of the $w_0 w_a \mathrm{CDM}$ model from DESI DR1.}
    \label{fig:w_evo}
\end{figure}

In particular, despite the small number of DESI BAO data points (only 6), the SdSDE model still produces a substantially larger $\chi^2_\mathrm{BAO}$. In figure \ref{fig:w_evo}, we show the evolution of the dark energy equation of state $w(a)$. The red dashed line corresponds to the SdSDE prediction using the best fit parameters in Table~\ref{tab_results} and the blue shaded regions indicate the 68\% and 95\% confidence level of the DESI DR1 official $w_0w_a \mathrm{CDM}$ constraints \cite{2025JCAP...02..021A}, using the publicly available data at \url{https://data.desi.lbl.gov/public/dr1/vac/dr1/bao-cosmo-params/v1.0/cobaya/base_w_wa/}). As shown in the figure, the behavior of $w(a)$ in the SdSDE model deviates significantly from the DESI best‐fit $w(a)$ at both low redshift ($z < 0.5$) and high redshift ($z > 3$). We conclude that the different low‐redshift evolution of $w(a)$ leads to a larger $\chi^2$ than expected, even though current DESI BAO constraints are still limited at higher redshift.

\begin{table}[ht]
    \centering
    \renewcommand{\arraystretch}{1.5}
    \begin{tabular}{|c|c|c|c|c|}
    \hline
    & $\mathrm{SdSDE}$ & $\Lambda \mathrm{CDM}$ \\ \hline 
    $\chi^2$ & $2884.12 \left( 2871.13 \right) ^{+3.65}_{-7.60}$ & $2836.12 \left( 2819.00 \right) ^{+1.63}_{-9.53}$ \\
    $\chi^2_\mathrm{CMB}$ & $2796.49 \left( 2783.92 \right) ^{+7.78}_{-3.89}$ & $2794.59 \left( 2777.11 \right) ^{+1.90}_{-9.69}$ \\
    $\chi^2_\mathrm{BAO}$ & $33.55 \left( 33.01 \right) ^{+0.50}_{-1.22}$ & $13.23 \left( 13.90 \right) ^{+0.95}_{-3.02}$ \\
    $\chi^2_\mathrm{SN}$ & $54.07 \left( 54.21 \right) ^{+1.90}_{-1.91}$ & $28.30 \left( 27.99 \right) ^{+0.58}_{-0.66}$ \\ \hline
    \end{tabular}
    \caption{The $\chi^2$ values of each likelihood. $\chi^2_\mathrm{CMB}$, $\chi^2_\mathrm{BAO}$, and $\chi^2_\mathrm{SN}$ correspond to the $\chi^2$ of  Planck+lensing, DESI DR2 BAO, and supernovae, respectively.}
    \label{tab_params4}
\end{table}

Finally, we discuss the implications for the Hubble constant. In the SdSDE model, the inferred $H_0$ is significantly higher than in the $\Lambda$CDM model. This aligns with earlier discussions in Ref. \cite{Croker:2024jfg}. Local measurements of $H_0$, such as from the SH0ES Collaboration, which reports $H_0 = 73.04 \pm 1.04 \, \mathrm{km/s/Mpc}$ \cite{2022ApJ...934L...7R}, have persistently exceeded the value inferred from Planck CMB observations under $\Lambda$CDM, a discrepancy known as the “Hubble tension” \cite{2020PhRvD.101d3533K, 2023Univ....9..393V}. Although including local $H_0$ measurements slightly improves the fit of the SdSDE model, the overall statistical preference for $\Lambda$CDM remains largely unchanged, and SdSDE continues to be strongly disfavored.

\section{Summary}
\label{sec:summary}

In this paper, we examined the concept of a dark energy model induced by the de-Sitter regions around black holes (SdSDE). We investigated whether SdSDE can serve as the origin of dark energy by examining their impact on cosmological observations. The cosmological data analysis for the SdSDE model was performed using Planck 2018 CMB data, as well as CMB lensing, DESI DR2 BAO, and Union3 supernovae datasets. In this study, the modification of black hole geometry was translated into an effective mass, and the energy density of SdSDEs was parameterized in terms of the conventional black hole mass density, as described in Eq.\ref{rho_ccbh}. Additionally, the relic black hole mass density estimated from the simulated mass function in Refs.\cite{2022ApJ...924...56S, 2022ApJ...934...66S} can be used in this context.

According to the MCMC analysis using the aforementioned datasets, several cosmological parameters showed significant deviations from their values in the $\Lambda$CDM model. In particular, we found a larger value of the Hubble constant $H_0$. However, the overall $\chi^2$ values obtained for the SdSDE model were worse than those for $\Lambda$CDM, and we found that the Akaike Information Criterion defference is $\Delta_\mathrm{AIC} = \Delta \chi^2 \sim 53$ since the number of free parameters is identical between the SdSDE and $\Lambda \mathrm{CDM}$ models. We therefore conclude that the SdSDE model is strongly disfavored by the combined CMB+BAO+SN data.

It should be noted that the current results are based on several approximations. These include the use of an approximate form for the mass function in Eq.~\ref{schechtergaussian} and the polynomial fitting of the black hole mass density in Eq.~\ref{rho_polynomial}. Recent JWST results indicate a larger number of AGNs than expected at high redshift. Therefore, the realistic black hole mass function may differ from the approximation used in this study.

Finally, while the contribution of black holes as SdSDEs is significant on the $\mathrm{Mpc}^3$ scale, the effect of SdS region is suppressed by a factor of approximately $10^{-16}$ compared to the conventional mass contribution on the $\mathrm{pc}^3$ scale. Therefore, it is difficult to probe the SdSDE model through astrophysical phenomena at small (galactic or binary) scales.

However, there is another possibility to constrain this SdSDE model. This model can be generalized as one in which a singularity generates a cosmological constant, implying that the constant emerges at the initial time of the universe. Hence, methodologies developed to discriminate among inflation models may be applicable to this cosmological coupling, but a detailed investigation is left for future work.

\section{Acknowledgement}
I thank Koichiro Nakashima and Genki Naruse for meaningful discussions. I also thank Prof. Kiyotomo Ichiki and Prof. Hironao Miyatake for their helpful comments on revising this paper.

\clearpage 
\bibliography{CCBH}

\end{document}